\documentclass[prb,showpacs,twocolumn,nobibnotes,epsf,sort&compress,super]{revtex4}

\usepackage{graphicx}
\usepackage{dcolumn}
\usepackage{bm}
\usepackage{SIunits}

\begin{document}
\title{Study on the oxygen isotope effect in A-site ordered manganite $R$BaMn$_{2}$O$_{6}$ ($R$ = La, Pr, Nd, Sm)}
\author{G. Y. Wang, K. H. Wong}
\author{Y. Wang}
\altaffiliation{Corresponding author}
\email{apywang@inet.polyu.edu.hk} \affiliation{Department of
Applied Physics and Materials Research Center, the Hong Kong
Polytechnic University, Hong Kong SAR, China\\}
\date{\today}

\begin{abstract}

A-site ordered manganites $R$BaMn$_{2}$O$_{6}$ ($R$ = La, Pr, Nd,
Sm) were synthesized and the oxygen isotope effect in them was
studied. It was found that the substitution of $^{16}$O by $^{18}$O
has caused increases in both the Neel temperature ($T$$_{N}$) and
the charge ordering temperature ($T$$_{CO}$) and an decrease in the
Curie temperature ($T$$_{C}$). The isotope exponent is small on
$T$$_{C}$ and $T$$_{CO}$ compared with that in A-site disordered
manganites such as La$_{1-x}$Ca$_{x}$MnO$_{3}$, which may be caused
by weaker electron-phonon coupling in A-site ordered ones. But the
isotope exponent of $T$$_{N}$ is large for $R$ = La. We didn't
observe the large oxygen-isotope shift that occurs in
NdBaMn$_{2}$O$_{6}$, where a multicritical point is believed to
appear near $R$ = Nd, and strong fluctuation due to the competition
between ferromagnetic metal and charge ordering phase should be
present.

\end{abstract}


\vskip 300 pt

\maketitle

\section{INTRODUCTION}

Perovskite-related manganites, e.g. $R$$_{1-x}$$A$$_{x}$MnO$_{3}$
($R$ = Y and rare earth metals, $A$ = Ca and Sr), have been
studied extensively since the colossal magnetoresistance (CMR)
phenomenon was first observed in
La$_{0.66}$Ca$_{0.33}$MnO$_{3}$\cite{Jin}. Complex phase diagram
has been observed in them, where ferromagnetic metal (FM),
antiferromagnetic (AF) insulator, charge ordering (CO) state, and
orbital ordering (OO) state present under different conditions.
The evolution of magnetic and electronic properties in manganites
have been investigated by changing parameters including
temperature, magnetic field, pressure, hole doping level, size of
A-site atom, doping on B-site, the quenched disorder, mismatch
effect, and so on.\cite{Dagotto2} It is now widely accepted that
the CMR effect is based on the phase competition between FM state
and insulating state in nanometer length scale, and will be much
enhanced near a bicritical point. A small bandwidth can also
enhance the CMR effect, and quenched disorder will expand the
parameter space where this effect can be observed\cite{Dagotto}.

It is known that the radii of the cations are a critical factor to
determine the crystal structure and properties of the perovskite
mangnites. For example, calcium and strontium ions, with the radii
$r$$_{Ca^{2+}}$ = 1.34 \AA$ $ and $r_{Sr^{2+}}$ = 1.44 \AA,
respectively, can readily substitute La$^{3+}$ ($r$ = 1.36 \AA)
because of their radii are similar. However, Ba ($r$ = 1.61 \AA) has
a rather low solubility in the system due to the much larger radius.
Especially for half-doping level, a novel structure
$R$BaMn$_{2}$O$_{6}$ with ordered A-site ions is obtained with
special treatment (depending on the difference of ion radii between
$R$$^{3+}$ and Ba$^{2+}$).\cite{Millange, Arima, Anthony, Akahoshi1,
Trukhanov, Nakajima1, Nakajima2, Nakajima3, Akahoshi2} In A-site
ordered manganites, the $R$-O and Ba-O layers alternatively stack
along the $c$-axis, doubling the lattice constant $c$(2$a$$_{P}$,
where $a$$_{P}$ is the axis of the simple perovskite cell). The
A-site ordered manganites $R$BaMn$_{2}$O$_{6}$ exhibit different
properties from the A-site disordered ones
$R$$_{0.5}$Ba$_{0.5}$MnO$_{3}$, such as higher Curie temperature
($T$$_{C}$) and charge ordering temperature ($T$$_{CO}$), and the
disappearance of spin glass (SG) state\cite{Millange, Nakajima3,
Akahoshi1}. They are divided into three different groups according
to their properties: (1). $R$ = Y and Ho-Tb, a structure phase
transition at $T$$_{t}$, a CE-type charge/orbital ordering (CO/OO)
transition at $T$$_{CO}$, and an AF transition at $T$$_{N}$; (2).
$R$ = Gd-Sm, a CE-type CO/OO transition at $T$$_{CO1}$, an AF
transition at $T$$_{N}$, and another CO/OO transition at $T$$_{CO2}$
with stack pattern different from the former one; and (3). $R$ =
Nd-La, a ferromagnetic transition at $T$$_{C}$ and an AF transition
at $T$$_{N}$. A multicritical point is also observed near $R$ = Nd,
where three ordered states, i.e., FM, A-type AF, and CO/OO states,
compete with each other\cite{Akahoshi1}. However, the attempt to
find CMR effect in this region is in vain, and large
magnetoresistance can only be found when quenched disorder is
introduced back into this system\cite{Ueda2}.

Oxygen isotope exchange is an useful tool to study the
electron-phonon coupling in materials, because it can avoid the
change in chemical environment and physical structure due to other
doping type, and introduce only small change in oxygen mass and
hence the energy of phonon.\cite{Zhao, Wang} Giant oxygen isotope
effect has been found in CMR materials such as
La$_{0.8}$Ca$_{0.2}$MnO$_{3}$ (21 K shift in $T$$_{C}$), indicating
the presence of small polarons in it\cite{Zhao}. In this paper, we
have studied the oxygen isotope effect in A-site ordered manganites
$R$BaMn$_{2}$O$_{6}$ ($R$ = La, Pr, Nd, Sm), aiming to establish a
physical picture of the electron-phonon coupling in them and the
relationship between quenched disorder and CMR effect. It was found
that the oxygen isotope effect on $T$$_{C}$ and $T$$_{CO}$ is small,
which may be caused by the absence of A-site disorder. But the
oxygen isotope effect on $T$$_{N}$ of LaBaMn$_{2}$O$_{6}$ is large,
which should be attributed to the fragileness of the AF transition
in this composition.

\section{EXPERIMENT DETAILS}
Ceramic samples of A-site ordered $R$BaMn$_{2}$O$_{6}$ ($R$ = La,
Pr, Nd, and Sm) were prepared by solid-state reaction method with
stoichiometric $R$$_{2}$O$_{3}$ (except for $R$ = Pr with
Pr$_{6}$O$_{11}$), BaCO$_{3}$, and MnO$_{2}$, which has been
reported everywhere\cite{Millange, Akahoshi2}. But for the following
oxygen isotope exchange, a process different from previous
one\cite{Wang} was employed, due to the sensitivity of A-site
ordering to the heat treatment and oxygen pressure. A ceramic sample
was cut into two pieces, and sealed into two quatz tubes (one with
$^{18}$O$_{2}$ and another with $^{16}$O$_{2}$) of a tube furnace.
The samples were first heated to 900 $\celsius$ for 48 h to ensure
the oxygen isotope exchange. It should be noticed that the A-site
order maybe partly destroyed after this heat treatment, as reported
in literature\cite{Nakajima2, Nakajima3}. After the samples were
cooled to room temperature, the oxygen gas was pumped out and some
hydrogen gas was filled into the tubes. Then the samples were heated
to 950 $\celsius$ for 5 hours, to reduce the sample and reconstruct
the A-site ordered structure. After the treatment the system was
cooled to 350 $\celsius$, then the tubes were evacuated again in
order to remove the residual hydrogen and water (formed during the
high temperature treatment). Subsequently oxygen gas was filled into
the two tubes, one with $^{18}$O$_{2}$ and another with
$^{16}$O$_{2}$. The samples were kept at 350 $\celsius$ for 24 h to
oxidize $R$BaMn$_{2}$O$_{5}$ to $R$BaMn$_{2}$O$_{6}$. The oxygen
isotope enrichment is determined by the weight change of samples.
The $^{18}$O samples have about 80($\pm$5)\% $^{18}$O and 20($\pm$
5)\% $^{16}$O. The magnetization of samples were measured by
vibrating sample magnetometer (Lakeshore 7400) from 80K to 400K. The
resistivity were measured in cryostat with a helium compressor by
four probe method. To reduce the measurement error, all the data
were recorded during the warming process. To confirm the oxygen
isotope enrichment, Raman spectra were recorded at room temperature
by a Horiba HR800 micro-Raman system, with a 488 nm laser as the
excitation source.

\section{RESULT AND DISCUSSION}

\begin{figure}[ht]
\centering
\includegraphics[width=8cm]{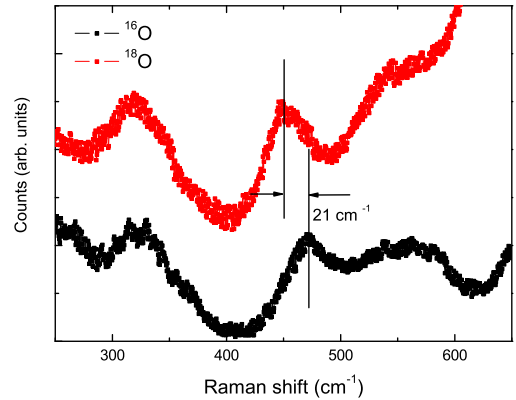}
\caption{(Color online) Raman spectra at room temperature for the
$^{16}$O and $^{18}$O samples of LaBaMn$_{2}$O$_{6}$,
respectively. The shift of Raman peak has been marked by two
vertical lines, from which the 76\% oxygen isotope enrichment can
be deduced. } \label{fig1}
\end{figure}

X-ray diffraction was performed on $^{16}$O and $^{18}$O samples of
$R$BaMn$_{2}$O$_{6}$ ($R$ = La, Pr, Nd, and Sm) to check the phase
purity and the presence of A-site ordering. No impurity phase was
discernible in the x-ray diffraction profiles. The signature of
A-site ordering, i.e. the (001) peak with $c$$\sim$2$a$$_{P}$, was
observed in both kinds of samples (data not shown here), which
confirmed the effectiveness of hydrogen annealing on the
reconstruction of A-site ordering. The Raman spectra of samples were
also recorded for the calculation of oxygen isotope enrichment. Fig.
1 shows the Raman spectra for the $^{16}$O and $^{18}$O samples of
LaBaMn$_{2}$O$_{6}$. A peak located at 471 cm$^{-1}$ was shifted to
450 cm$^{-1}$ after the oxygen isotope substitution. The frequency
of Raman mode $f$ is reciprocally proportion to the square root of
oxygen mass: $f$($^{18}O$)/$f$($^{16}O$) = $\sqrt{16/M'}$. The
oxygen isotope enrichment deduced from Raman shift is 76\%,
consistent with that deduced from the mass increase (80$\pm$5\%).

\begin{figure}[t]
\centering
\includegraphics[width=8cm]{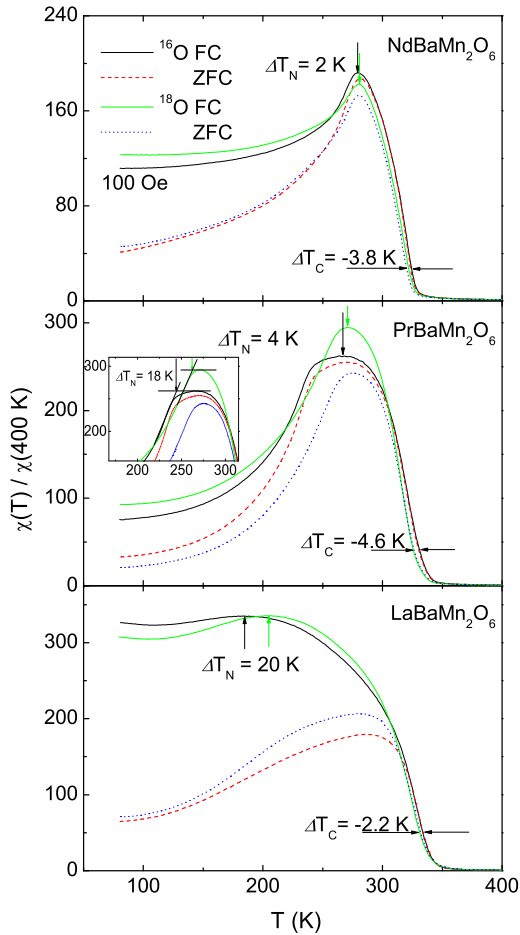}
\caption{(Color online) Temperature dependence of magnetization
for the $^{16}$O and $^{18}$O samples of $R$BaMn$_{2}$O$_{6}$ ($R$
= La, Pr, and Nd) at H = 100 Oe. The shift of T$_{C}$ and T$_{N}$
has been marked by horizontal arrows and vertical arrows,
respectively. Inset shows another way to define the shift of
$T$$_{N}$, which give a larger value.} \label{fig2}
\end{figure}

The temperature dependence of magnetization (normalized to the data
at 400K) for the $^{16}$O and $^{18}$O samples of
$R$BaMn$_{2}$O$_{6}$ ($R$ = La, Pr, and Nd) are shown in Fig. 2. All
the data were recorded during the warming cycle for field cooling
(FC) under a field of 100 Oe and zero field cooling (ZFC)
situations. Ferromagnetic order is observed in all three samples.
After the substitution of $^{16}$O by $^{18}$O, $T$$_{C}$ is shifted
to a lower temperature, marked by two horizontal arrows in Fig. 2.
The shifts of $T$$_{C}$ are -2.2 K, -4.6 K, and -3.8 K for $R$ = La,
Pr, and Nd, respectively. Comparing with the $T$$_{C}$ shift in
A-site disordered samples, for instance, $\Delta$$T$$_{C}$ = 21 K
for La$_{0.8}$Ca$_{0.2}$MnO$_{3}$\cite{Zhao}, and even a
metal-insulator transition caused by oxygen isotope
exchange,\cite{Babushikina} the current ones are much smaller. With
decreasing temperature, a decrease in FC magnetization can be found
in all the three $^{16}$O samples, which is attributed to the
CE-type antiferromagnetic transition for $R$ = La, and A-type
antiferromagnetic transition for $R$ = Pr and Nd\cite{Nakajima2}.
Here, we regard $T$$_{N}$ as the temperature at which the maximum of
magnetization in FC curves is achieved, as indicated by the
perpendicular arrows in Fig.2. For the $^{18}$O samples, $T$$_{N}$
is shifted to a higher temperature, and the shifts are 20 K, 4 K,
and 2 K for $R$ = La, Pr and Nd, respectively. It should be pointed
out that the $T$$_{N}$ shift is strongly affected by the way how it
was defined. For example, we will get 18 K for $R$ = Pr, if we
determine the $T$$_{N}$ shift as the inset of Fig. 2. The shift of
$T$$_{N}$ is distinct for $R$ = La, but almost smeared out
completely for $R$ = Nd. This is very strange, since the critical
region is believed to located near $R$ = Nd, where fluctuation
should be enhanced due to the competition of three ordered FM,
A-type AFM and CO/OO states.

\begin{figure}[htp]
\centering
\includegraphics[width=8cm]{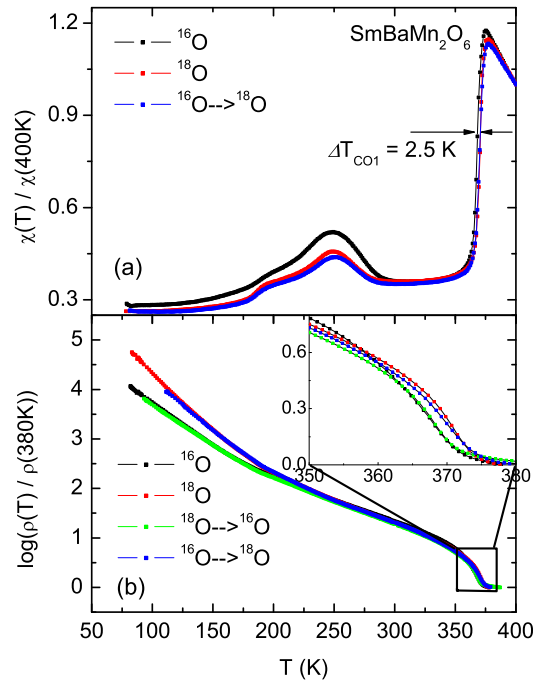}
\caption{(Color online) Temperature dependence of (a)
magnetization (normalized to 400 K) and (b) resistivity
(normalized to 380 K) for the $^{16}$O and $^{18}$O samples of
SmBaMn$_{2}$O$_{6}$. Some data from back-exchanged samples are
also presented. Inset of (b) shows the enlarged figure near the
area of the first CO/OO transition.} \label{fig3}
\end{figure}

Fig. 3 shows the temperature dependence of magnetization (normalized
to 400K) and resistivity (normalized to 380K) of
SmBaMn$_{2}$O$_{6}$. It should be noted that the magnetic field used
here is 1000 Oe. The shape of the magnetization is a little
different from that reported in literature\cite{Arima, Akahoshi1,
Nakajima2}, which may be caused by slight fluctuation in sample
composition among different groups. For the $^{16}$O sample, the
sharp decrease in the magnetization at 374 K ($T$$_{CO1}$) is
related to the first charge/orbital ordering transition\cite{Arima},
associated with an abrupt increasing of resistivity. When $^{16}$O
is replaced by $^{18}$O, this ordered temperature is shifted to a
higher temperature with $\Delta$$T$$_{CO1}$ = 2.5K, consistent with
that in resistivity shown in the inset of Fig. 3(b). At 248 K, a
peak is observed in the magnetization curve, which is attributed to
the CE-type AF transition\cite{Arima}. This transition, however, is
not reflected in the resistivity curve. The oxygen-isotope shift of
$T$$_{N}$ can only be seen in magnetization, from 248K to 250K. It
should be emphasized that the magnetization in the $^{18}$O sample
is smaller than that in the $^{16}$O sample. At about 193K, the
second CO/OO transition occurs, which shows a slight kink in
magnetization and a change of slope in resistivity. And the kink in
magnetization is clearer in the $^{18}$O sample. But the shift of
$T$$_{CO2}$ is difficult to calculate due to either the faintness of
this kink in the $^{16}$O sample, or the smallness of the
oxygen-isotope shift. Powder neutron diffraction has given the
evidence that the magnetic period in $c$-axis at T$<$T$_{CO2}$ is
4$a_{P}$, and deduced ABAB-type stacking pattern of Mn-O layer along
this direction\cite{Arima}. However, recent resonant soft x-ray
powder diffraction showed that the stacking is AAAA-type along
$c$-axis\cite{Garcia}. From the Goodenough-Kanamori rules, ABAB
stacking pattern will produce ferromagnetic coupling while AAAA
stacking pattern will produce AF coupling among the Mn-O layers in
the $c$ direction. The sudden drop in magnetization at $T$$_{CO2}$,
caused by the oxygen isotope substitution, may be regard as an
evidence for the AAAA-type stacking.

The oxygen isotope exponents for A-site ordered manganites
$R$BaMn$_{2}$O$_{6}$ ($R$ = La, Pr, Nd, and Sm) are shown in Fig.
4 (a), which are calculated using the formula: $\alpha$ =
-$d$ln$\textit{T}$/$d$ln$\textit{M}$ =
(16$\times$(\textit{T}$_{16}$-\textit{T}$_{18}$))/(\textit{T}$_{16}$$\times$(17.52-16)).
The oxygen isotope effects on the transition temperature are
presented in Fig. 4 (b), vs the radii of R$^{3+}$. Generally, the
substitution of $^{16}$O by $^{18}$O will soften the phonon, and
increase the effective mass of electrons by some degree of
electron-phonon coupling. Then the bandwidth of bare electron will
be reduced, which strengthens the insulating phase and weakens the
metal phase. This is the reason that the CO/OO ordering
temperature $T$$_{CO1}$, associated with a metal-insulator
transition in resistivity, is shifted to a higher temperature for
$R$ = Sm. The increased resistivity at low temperature in the
$^{18}$O sample is also an evidence for the reduced bandwidth. It
is believed that $T$$_{C}$ is proportional to the effective
bandwidth of electrons,\cite{Furukawa} so the substitution of
$^{16}$O by $^{18}$O will decrease the $T_{C}$, as we have
observed for $R$ = La, Pr and Nd. On the other hand, decreasing
bandwidth means reduced hoping integral $t$, which will favor the
AF interaction and suppress the ferromagnetic interaction between
Mn ions. So the $T_{N}$ is shifted to a higher temperature for $R$
= La, Pr, Nd, in spite of the transition type A or
CE\cite{Nakajima2}.

\begin{figure}[t]
\centering
\includegraphics[width=8cm]{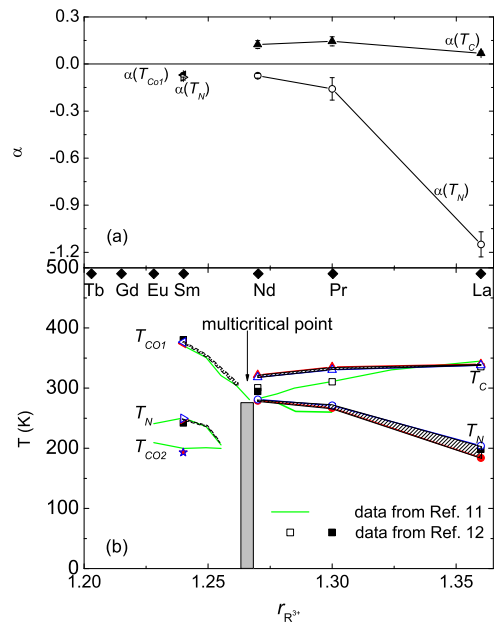}
\caption{(Color online) Part of the phase diagram for
$R$BaMn$_{2}$O$_{6}$ ($R$ = La, Pr, and Nd). Green line are the
data taken from Ref. 11, square data are taken from Ref. 12. Red
solid points and blue hollow points are for our $^{16}$O and
$^{18}$O samples, respectively. Hatched areas indicate the
possible change caused by oxygen isotope exchange.} \label{fig4}
\end{figure}

We then further consider the magnitudes of the oxygen-isotope
shifts. First, all the shifts of $T$$_{C}$ and $T$$_{CO}$ are
small compared with that in A-site disordered ones. Quenched
disorder will reduce the bandwidth\cite{Motome} (this is the
reason that the curie temperature in ordered LaBaMn$_{2}$O$_{6}$
is 50 K higher than that in disordered
La$_{0.5}$Ba$_{0.5}$MnO$_{3}$), which enhanced the relative
importance of electron-phonon coupling\cite{Dagotto2}, or in other
word, the oxygen isotope shift of $T$$_{C}$. For $R$ = Sm, the
absence of disorder will also make the CO phase tough, which
suppress the relative importance of electron-phonon coupling, and
give small oxygen isotope shift. So the small oxygen isotope shift
of $T$$_{C}$ and $T$$_{CO}$ in ordered $R$BaMn$_{2}$O$_{6}$ is
accessible. And the $T$$_{C}$ shift for $R$ = La is the smallest,
due to the highest $T$$_{C}$ (or largest bandwidth) in it. In
fact, small oxygen isotope effect not only observed in A-site
ordered manganites here, but also observed in other A-site ordered
perovskite materials such as $R$BaCo$_{2}$O$_{5+x}$\cite{Conder}.
One may expect larger oxygen-isotope shift on $T$$_{C}$ in
Nd$_{1-x}$Sm$_{x}$Mn$_{2}$O$_{6}$, since it seems that the
$r$$_{R^{3+}}$ of the critical point should be sligntly smaller
than $r$$_{Nd^{3+}}$ in our experiment.

Second, the isotope shift of $T$$_{N}$ is large for $R$ = La, but
small for $R$ = Nd. Neutron powder diffraction experiment shows that
the FM transition in $R$BaMn$_{2}$O$_{6}$ is a second-ordered one,
while the AF transition is a first-ordered one\cite{Nakajima2,Sato}.
It is obvious that the lattice will couple more strongly to the
first-ordered transition. This is the reason that we observed a
large oxygen-isotope shift in $T$$_{N}$ but not in $T$$_{C}$ for $R$
= La. The AF transition is especially fragile for $R$ = La, and the
nuclear magnetic resonant experiment demonstrated that AF phase
occupies only about half the volume of the whole
sample\cite{Kawasaki}. The fragility of T$_{N}$ will also enhance
the relative importance of electron-phonon coupling on it, and
induces the largest oxygen isotope shift in our experiment. But for
$R$ = Nd, the case is different. The multicritical point is believed
to locate near $R$ = Nd, where enhanced fluctuation should be
present due to the competition among FM, CO/OO phase, and A-type AF
phase. In A-site disordered samples such as ($R$,Ca)MnO$_{3}$ and
($R$,Sr)MnO$_{3}$, the critical point is always related to the
enhanced fluctuation due to the competition between FM phase and CO
insulating phase, and the CMR effect\cite{Dagotto2,Sen}. A large
oxygen-isotope shift on $T$$_{N}$ is also expected for $R$ = Nd.
However, $T$$_{N}$ in NdBaMn$_{2}$O$_{6}$ is almost the highest in
the phase diagram of $R$BaMn$_{2}$O$_{6}$\cite{Akahoshi1, Ueda1},
indicating the relatively strong AF phase in it. The firmness of AF
phase will certainly reduce the relative importance of oxygen
isotope substitution, and show limited response on it.

\section{CONCLUSION}

In summary, oxygen isotope effect on the FM, AF transition, and
CO/OO transition is studied for $R$BaMn$_{2}$O$_{6}$ ($R$ = La,
Pr, and Nd). Small oxygen isotope shift is observed on $T$$_{CO1}$
for $R$ = Sm, due to the tough CO state caused by the absence of
A-site disorder. The first-ordered nature of AF phase for $R$ = La
produces much larger oxygen isotope shift on $T$$_{N}$ than on the
second-ordered FM transition temperature $T$$_{C}$. For $R$ = Nd,
the isotope effect on $T$$_{C}$ is slightly enhanced due to the
reduced band width comparing with that of $R$ = La; but the oxygen
isotope shift on T$_{N}$ is smaller due to the relative strong AF
phase in it.

\section*{ACKNOWLEDGMENTS} This work was supported by the Hong Kong
Polytechnic University Postdoctoral Scheme (G-YX0C). The support
from the Center for Smart Materials is also acknowledged.\\

\end{document}